\documentclass[aps, pra, reprint, superscriptaddress, showkeys]{revtex4-1}
\usepackage[english]{babel}
\usepackage{times}
\usepackage{w-thm}
\usepackage{amsfonts,amsmath,amssymb}
\usepackage[margin=25mm,inner=20mm,marginpar=10mm]{geometry}
\usepackage{graphicx}

\begin{document}
\noindent
%\DOIsuffix{theDOIsuffix}
%%
%% issueinfo for header and copyright line
%\Volume{16}
%\Issue{1}
%\Copyrightissue{01}
%\Month{01}
%\Year{2015}
%\pagespan{1}{}
%%
%\Receiveddate{12 March 2015}
%\Reviseddate{30 November 2007}
%\Accepteddate{2 December 2007}
%\Dateposted{3 December 2007}
%\keywords{Matter waves, Quantum optics, Casimir--Polder forces, van der Waals forces} 

\title{An atomically thin matter-wave beamsplitter}

\author{Christian Brand}
\affiliation{University of Vienna, Faculty of Physics, VCQ, QuNaBioS, Boltzmanngasse 5, A-1090 Vienna, Austria}

\author{Michele Sclafani}
\affiliation{University of Vienna, Faculty of Physics, VCQ, QuNaBioS, Boltzmanngasse 5, A-1090 Vienna, Austria}
\affiliation{ICFO - Institut de Ci\`ences Fot\`oniques, 08860 Castelldefels (Barcelona), Spain}

\author{Christian Knobloch}
\affiliation{University of Vienna, Faculty of Physics, VCQ, QuNaBioS, Boltzmanngasse 5, A-1090 Vienna, Austria}

\author{Yigal Lilach}
\affiliation{The Center for Nanosciences and Nanotechnology at Tel Aviv University, Tel Aviv 69978, Israel}

\author{Thomas Juffmann}
\affiliation{University of Vienna, Faculty of Physics, VCQ, QuNaBioS, Boltzmanngasse 5, A-1090 Vienna, Austria}
\affiliation{Stanford University, Physics Department, 382 Via Pueblo Mall, Stanford, CA 94305-4060, USA}

\author{Jani Kotakoski}
\affiliation{University of Vienna, Faculty of Physics, PNM, Boltzmanngasse 5, A-1090 Vienna, Austria}

\author{Clemens Mangler}
\affiliation{University of Vienna, Faculty of Physics, PNM, Boltzmanngasse 5, A-1090 Vienna, Austria}

\author{Andreas Winter}
\affiliation{Friedrich Schiller University Jena, Institute of Physical Chemistry, Lessingstr. 10, D-07743 Jena, Germany}

\author{Andrey Turchanin}
\affiliation{Friedrich Schiller University Jena, Institute of Physical Chemistry, Lessingstr. 10, D-07743 Jena, Germany}

\author{Jannik Meyer}
\affiliation{University of Vienna, Faculty of Physics, PNM, Boltzmanngasse 5, A-1090 Vienna, Austria}

\author{Ori Cheshnovsky}
\affiliation{The Center for Nanosciences and Nanotechnology and School of Chemistry, The Raymond and Beverly Faculty of Exact Sciences, Tel Aviv University, Tel Aviv 69978, Israel}

\author{Markus Arndt}
\thanks{E-mail:~\textsf{markus.arndt@univie.ac.at}, Phone: +00\,43\,1\,4277\,51210, Fax: +00\,43\,1\,4277\,9512}
\affiliation{University of Vienna, Faculty of Physics, VCQ, QuNaBioS, Boltzmanngasse 5, A-1090 Vienna, Austria}

\begin{abstract}
Matter-wave interferometry has become an essential tool in studies on the foundations of quantum physics \cite{Arndt2014} and for precision measurements \cite{Cronin2009, Rosi2014, Bouchendira2011, Dickerson2013, Geiger2011}. Mechanical gratings have played an important role as coherent beamsplitters for atoms \cite{Keith1988}, molecules and clusters \cite{Schollkopf1994, Arndt1999} since the basic diffraction mechanism is the same for all particles. However, polarizable objects may experience van der Waals shifts when they pass the grating walls \cite{Grisenti1999, Lonij2010} and the undesired dephasing may prevent interferometry with massive objects \cite{Gerlich2007}. 
Here we explore how to minimize this perturbation by reducing the thickness of the diffraction mask to its ultimate physical limit, i.e. the thickness of a single atom. We have fabricated diffraction masks in single-layer and bilayer graphene as well as in $1$~nm thin carbonaceous biphenyl membrane. We identify conditions to transform an array of single layer graphene nanoribbons into a grating of carbon nanoscrolls. 
We show that all these ultra-thin nanomasks can be used for high-contrast quantum diffraction of massive molecules. They can be seen as a nanomechanical answer to the question debated by Bohr and Einstein \cite{Bohr1949} whether a softly suspended double slit would destroy quantum interference. In agreement with Bohr's reasoning we show that quantum coherence prevails even in the limit of atomically thin gratings.
 \end{abstract}
\maketitle

\section{Introduction}
The key to all matter-wave interferometry are coherent beamsplitters which divide each incident wave into separated wavelets with well-defined phase relations \cite{Cronin2009, Juffmann2013}. While modern atom interferometry often utilizes the momentum recoil of resonant laser light to split the atomic wave \cite{Borde1989, Kasevich1991}, mechanical masks \cite{Keith1988, Arndt1999} as well as optical phase \cite{Gould1986} or absorption \cite{Moskowitz1983} gratings can also be used to divide a matter-wave front. Since the universality of mechanical gratings is compromised by the van der Waals (vdW) potential it is important to ask to what extent it is possible to minimize this interaction by reducing the grating thickness. This technological feat - to actually create free-standing nanogratings in even a single layer of atoms \cite{Geim2007} - is accompanied by the fundamental question whether the path of a massive particle through the multi-slit array will become significantly entangled with the mechanical motion of the recoiling ultra-thin structure. If this were the case we would expect to observe loss of the interference contrast.\\

\section{Results and Discussion}
 
Using a focused ion beam we were able to mill a periodic diffraction grating into a single layer of graphene that was suspended over a silicon nitride (SiN$_\text{x}$) membrane (Fig. \ref{fig:gratings}a, see Appendix Sec.\ref{app:gratings} for details). The nanoribbons ($64\pm3$~nm wide) were written with a period of $88\pm3$~nm (Fig. \ref{fig:gratings}c). They spontaneously transform into carbon nanoscrolls \cite{Lucot2009}, here with a diameter down to $8$~nm (Fig. 1b and Appendix Sec.\ref{app:scrolls}). For shorter grating bars scanning transmission electron microscopy (STEM) reveals, however, stable flat single-layer graphene structures (Fig. \ref{fig:gratings}e). 
We were also able to write gratings into bilayer graphene suspended across a lacey carbon mesh (Fig. \ref{fig:gratings}f). The second layer of carbon atoms suppresses the formation of nanoscrolls entirely on the observed scale. Bilayer graphene may also form bonds at open cuts and thus expose closed edges and a thicker wall than expected based on the number of layers alone \cite{Liu2009}. We compare these structures finally to an insulating structure of almost identical thickness, the carbonaceous biphenyl membrane \cite{Angelova2013} (also on lacey carbon, Fig. \ref{fig:gratings}g,) and reference all images to the diffraction of molecules at $45$~nm thick silicon nitride (Fig. \ref{fig:gratings}h) \cite{Juffmann2012}.

\begin{figure*}[htb]
\begin{center}
\includegraphics[width=16cm]{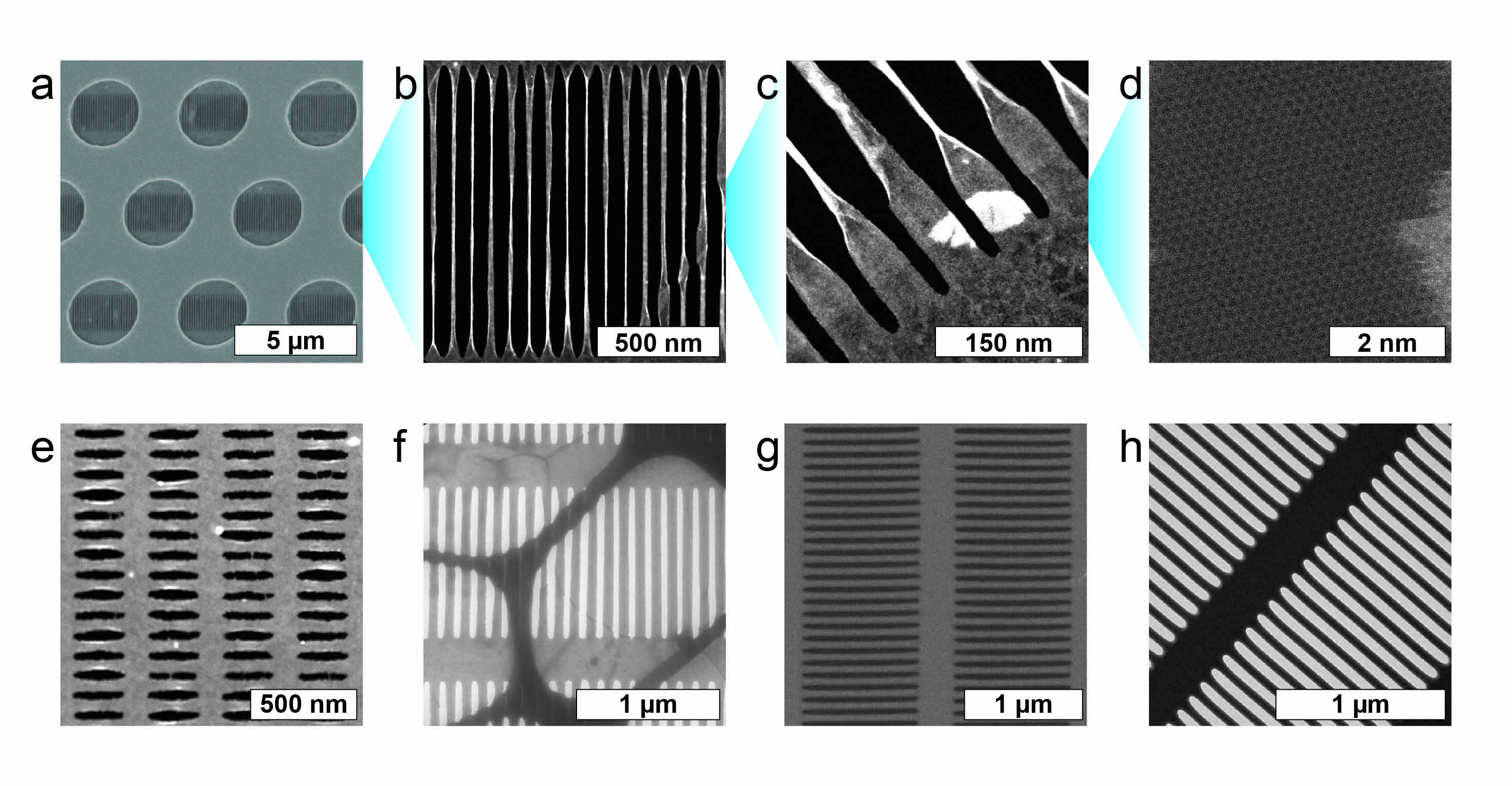}
\end{center}
\caption{Exploring the ultimate limit of nanomechanical diffraction gratings. (a) Gratings written into single-layer graphene suspended across a holey SiN$_\text{x}$ membrane. (b) Carbon nanoscrolls spontaneously form, when the aspect ratio of the graphene nanoribbons becomes too high (b) and (c). Folds and contaminations can prevent the membrane from curling. (d) STEM visualizes the hexagonal atomic structure of single-layer graphene. (e) Reducing the length of the $40$~nm wide nanoribbons to $250$~nm prevents the curling. (f) Adding another layer of carbon atoms (bilayer graphene) allows us to substantially extend the aspect ratio and to keep the grating flat. (g) Large area gratings can also be written into a carbonaceous biphenyl membrane. (h) Grating in silicon nitride of $45$~nm thickness}
\label{fig:gratings}
\end{figure*}

\begin{figure*}[htb]
\begin{center}
\includegraphics[width=16cm]{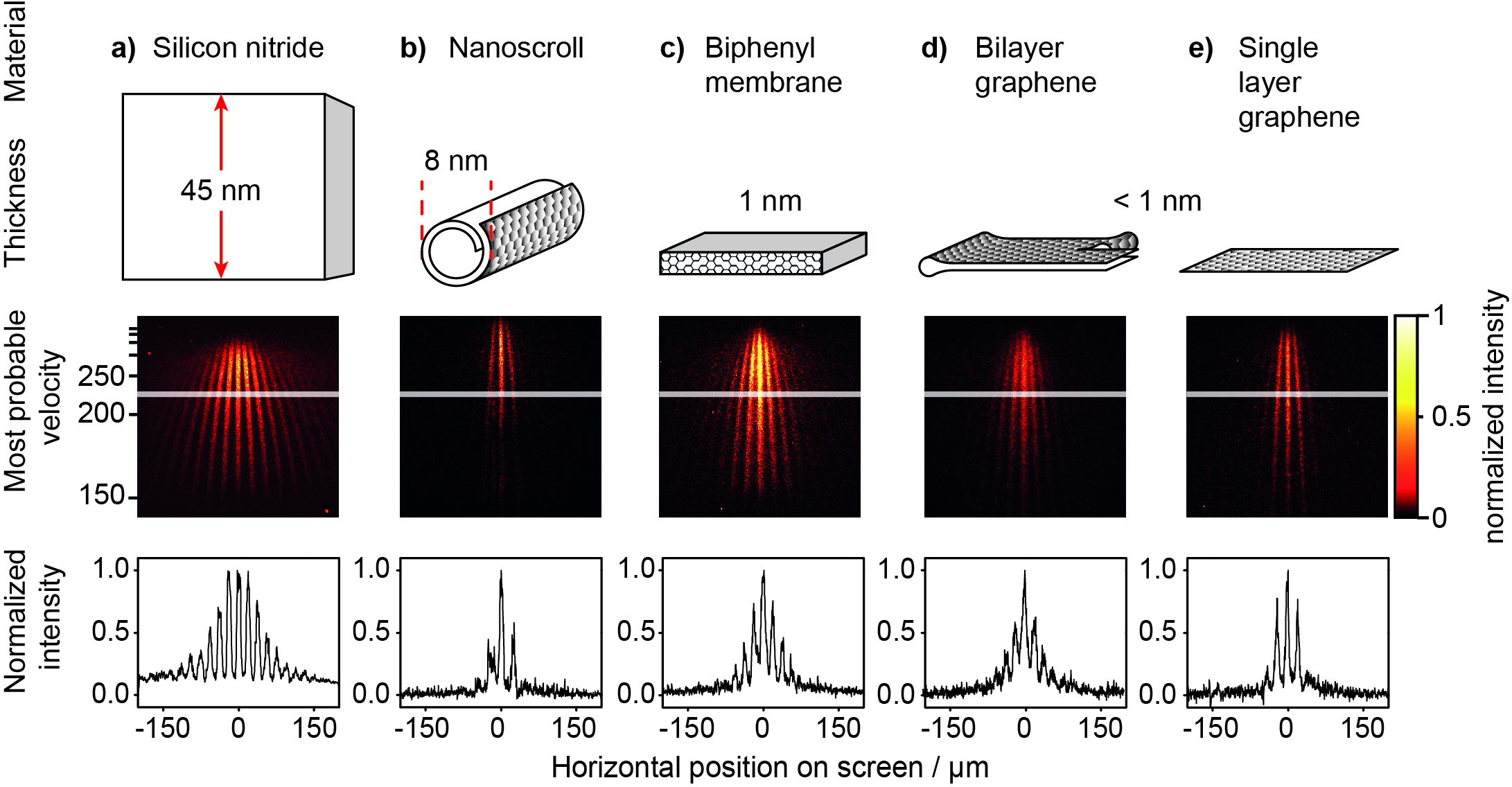}
\end{center}
\caption{Influence of the membrane thickness and material composition of the van der Waals interaction in molecule diffraction. Top row: Grating type and thickness. (a) Period $d=105\pm1$~nm, (b) $d=88\pm3$~nm, (c) $d=107\pm9$~nm, (d) $d=113\pm4$~nm, (e) $d=101\pm2$~nm. Middle row: Fluorescence image of the PcH$_2$ diffraction image. With increasing flight time the molecules fall deeper in the gravitational field. Fast molecules arrive at the top, slow molecules arrive at the bottom of the detector. Bottom row: Each trace represents a normalized vertical 1D integral over the area under the white line, corresponding to the molecular velocity band centered at $220$~m/s. The experimental data can be approximated by Kirchhoff-Fresnel diffraction theory (see Appendix \ref{app:Kirchhoff}) by taking into account that  the van der Waals forces reduce the effective slit width \cite{Grisenti1999}:  (a) geometrical slit width $s= 50\pm2$~nm, effective slit width $s_\text{eff} = 15$~nm, (b) $s=65\pm6$~nm, $s_\text{eff} = 49$~nm, (c) $s=54\pm4$~nm, $s_\text{eff} = 28$~nm, (d) $s=62\pm8$~nm, $s_\text{eff} = 28$~nm, (e) $s=59\pm6$~nm, $s_\text{eff} = 35$~nm.}
\label{fig:overview}
\end{figure*}

\begin{figure}[htb]
\begin{center}
\includegraphics[width=7cm]{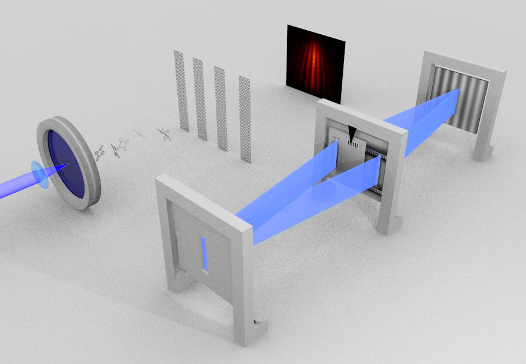}
\end{center}
\caption{A nanomechanical implementation of the Bohr-Einstein debate: Can atomically thin gratings be compatible with high-contrast interference? A sketch of the diffraction setup using ultra-thin gratings is depicted on the left, corresponding to Bohr's double slit thought experiment on the right \cite{Bohr1949}. In analogy to a debate between Bohr and Einstein we ask: Can a flexible suspension of the mechanical mount encode which-path information of particles in transition through the double slit?}
\label{fig:Born}
\end{figure}

Figure \ref{fig:gratings} demonstrates that stable structures can be written even into atomically thin membranes. The diffraction pattern behind a purely absorptive periodic mask - \emph{i.e.} without any phase modulation – can be described as the convolution of two contributions: the diffraction at each single slit of width $s$ and the diffraction at an array of infinitely thin slits of period $d$. The fringe positions are well described by wave theory. The fringe amplitudes are modulated by the single-slit diffraction pattern. 
Comparing Fig. \ref{fig:overview}(a) - (e), we see immediately vast differences in the diffraction at geometrically similar gratings. For $45$~nm thick silicon nitride (Fig. \ref{fig:overview}a) we observe interference up to the $9$th diffraction order, which can only be understood if we complement the quantum wave model by the assumption that the position-dependent phases accumulated in the presence of vdW forces alter the effective transmission function. Approximating this interaction by reducing the effective slit width \cite{Grisenti1999} (see Appendix Sec.\ref{app:Kirchhoff}) allows us to estimate the strength of the vdW interactions. For silicon nitride the analysis yields an effective slit width of $15$~nm - a reduction of the open width by $s/s_\text{eff} =3.3$. This influence is strongly reduced for the single-layer graphene grating, the probably thinnest conceivable grating (Fig. \ref{fig:overview}e, $s/s_\text{eff} = 1.7$). As expected this leads to a strong suppression of all diffraction orders beyond the first one. This holds true - and even more so - for gratings made from nanoscrolls ($s/s_\text{eff} = 1.3$), which maximize the opening fraction, \emph{i.e.} the ratio of slit width to period.
In contrast to that, the diffraction pattern both behind the carbonaceous biphenyl membrane ($s/s_\text{eff} =1.9$) and the bilayer graphene ($s/s_\text{eff} = 2.2$) show substantial populations up to the fifth diffraction order. This indicates a stronger influence of the attractive van der Waals force. While the carbonaceous biphenyl membrane should be an insulator, the bilayer sample is conducting. However, both imprint similar phase shifts onto the transmitted molecules. We attribute the remaining van der Waals forces in particular also to the lacey carbon support structure which is thicker (up to $100$~nm) and spatially less well controlled than the SiN$_\text{x}$ support of the other ultra-thin gratings.
The central result of our diffraction experiments is the observation of high-contrast interference for all gratings, where the contrast reaches its maximum when the interference minima approach the zero (background) level. It is the absence of signal in these minima which contradicts any classical expectation the most. Our result relates to a thought experiment between Bohr and Einstein who elucidated the role of decoherence in double slit diffraction \cite{Bohr1949}. Einstein predicted that the observer should be capable of extracting which-path information from the recoil that the double slit receives upon diffraction of a particle (Fig. \ref{fig:Born}). Invoking Heisenberg's uncertainty relation, Bohr was able to demonstrate, however, that the recoil imparted by the diffracted particles remains within the momentum uncertainty of the grating if this is well positioned.

Different recent studies have realized the probably smallest double slits and analogs to this Einstein-Bohr debate using diatomic molecules as the diffraction elements in a photoionization process \cite{Zimmermann2008, Akoury2007, Liu2015}. They were destroyed in the diffraction process and therefore were capable of carrying away path information. Our carbon nanostructures are probably the thinnest durable realization of the Bohr-Einstein idea. These atomically thin membranes are flexible but they last for months and survive the diffraction of millions of molecules. The observed full contrast shows that coherence prevails, which we interpret as an indication that the momentum exchange with the grating remains within the intrinsic momentum uncertainty of the grating bars (see Appendix \ref{app:Heisenberg}) \cite{Bohr1949}. Hence, diffraction does not lead to a loss of fringe visibility even though it can cause a sizable matter-wave (vdW) phase shift.
Graphene gratings are ten times thinner than any other beamsplitter for atoms, molecules or clusters before and they are four orders of magnitude thinner than the width of a typical laser grating \cite{Gerlich2007}. Our results on single-layer graphene are the best approach with regard to the diffraction of high-mass molecules at nanomechanical gratings, since they show that the effect of molecule-surface interactions can be substantially reduced, even though they do not eliminate them entirely. 
This property makes graphene nanogratings also appealing for use with other kinds of matter-waves. The reduced phase components may enable new coherence experiments with slow ions or anti-matter \cite{Hamilton2014}. It will also be interesting to explore atomically thin but insulating 2D sheets like boron nitride or molybdenum disulfide, in the future.
Even though van der Waals forces in thick gratings can be prohibitive for high-mass diffraction, they can also serve a purpose in metrology: our experiments show that a thick material grating (SiN$_x$, Fig. \ref{fig:overview}a) can act as a large momentum transfer beamsplitter. If we were to calibrate the momentum exchange between the extreme diffraction orders ($\pm9th$ order for $d=100$~nm) in units of the photon momentum ($\hbar k_\text{Rb}$) that is usually imparted in state-of-the-art beamsplitters \cite{Kasevich1991, Borde1989} for (rubidium) atoms, we see that the diffraction at our nanogratings amounts to $\Delta p=141\, \hbar k_\text{Rb}$. Photolithography can reliably provide thick gratings with a periodicity of better than $0.1$~nm over large areas \cite{Gerlich2007}. Stable thick membranes with narrow slits are predicted to provide a momentum transfer even beyond $500$~$\hbar k_\text{Rb}$. 

\section{Conclusion}

Finally, our work shows that path-decoherence close to material gratings is still negligible in all settings discussed here. This applies in particular for the Bohr-Einstein Gedankenexperiment. Even the conceivably thinnest durable mechanical grating is still sufficiently massive and sufficiently localized not to encode sizable recoil-information. The intrinsic momentum uncertainty of each grating bar is larger than the recoil imparted on each diffracted molecule (see Appendix \ref{app:Heisenberg}). 
This holds for all particles, independent of their mass, as long as we can keep their de Broglie wave-length, \emph{i.e.} their momentum, the same. For instance, quantum diffraction of insulin molecules seems conceivable once we can provide a directed molecular beam of about $20~$m/s velocity. Finally, it will also be intriguing to explore the role of quantum friction and path-decoherence \cite{Scheel2012, Anglin1997} which may become particularly relevant for polar molecules in the future.

\section*{Acknowledgements}
We acknowledge support by the European Commission (304886), the European Research Council (320694) and the Austrian science funds (DK CoQuS W1210-3). CB acknowledges the financial support of the Alexander von Humboldt foundation through a Feodor Lynen fellowship. JK acknowledges the Austrian Science Fund FWF for funding through project M 1481-N20. JM and CM acknowledge support from the Austrian science funds FWF project P 25721-N20. AW and AT acknowledge support of the DFG (SPP "Graphene" TU149/2-2, Heisenberg Program TU149/3-1). TJ acknowledges support by the Gordon and Betty Moore Foundation. We thank the group of Prof. Schattschneider, USTEM TU Vienna for assistance in recording image \ref{fig:gratings}(h). We thank Stefan Scheel and Johannes Fielder (Univ. Rostock) for fruitful discussions.

%\begin{thebibliography}{10}
%\bibliography{NatNano}	

%\end{thebibliography}

\section{Appendix}
\subsection{Limits imposed by the uncertainty principle}
\label{app:Heisenberg}

Coherent diffraction of a matter-wave is possible as long as the momentum transferred by the particle to the grating is within the momentum uncertainty of the grating itself. 	
In diffraction experiments at a single slit, the relationship $\Delta x \cdot \Delta p^\text{diff}\geq 0.89 h$ holds \cite{Nairz2002}, when $\Delta x=s$ designates the width of the single slit and  $\Delta p^\text{diff}$ the full width at half maximum (FWHM) of the diffraction curve. The momentum uncertainty $\Delta p^\text{diff}\geq 0.89 h/s$ is the FWHM of the envelope to all diffraction orders of the grating. This momentum is transferred to the grating, which itself is described by the uncertainty relationship $\Delta x \cdot \Delta p^\text{grat} \geq \hbar/2$. To estimate $\Delta p$ within the plane of the grating we need an independent measure of its position and movement at room temperature. 
For a freely suspended single-layer graphene ribbon (length $\times$ width = $2.8 \times 0.3-0.5$~$\mu$m) clamped on both sides capacity measurements by Garcia-Sanchez \emph{et al.} \cite{CarciaSanchez2008} allowed them to estimate an out-of-plane vibration amplitude of $\sigma \simeq 0.1$~nm. The vibration amplitude of shorter ribbons and in particular of in-plane modes are expected to be even smaller - due to their large width-to-thickness ratio. The value of $0.1$~nm therefore constitutes an upper boundary for the in-plane vibration amplitude for this system. 
The largest amplitudes for in-plane vibrations are expected for the nanoscrolls grating because of the (nearly) rotational symmetry of the scrolls. Their vibrational properties might be inferred from the results for carbon nanotubes. Here, the maximum displacement of a doubly clamped single-walled carbon nanotube due to thermal fluctuations is given as \cite{Sapmaz2003}:
\begin{equation}
 \sigma=\sqrt{\frac{k_b T L^3}{192 \cdot Y \cdot I}}
\end{equation}
where $L$ is the length of the nanotube, $Y=1$~TPa is the Young's modulus, here taken to be the value of a single wall carbon nanotubes, and $I=\pi\cdot d^4/64$ is the area moment of inertia, with $d$ the scroll diameter. For the geometry of our nanoscroll bars $(L=1.34$~$\mu$m, $d=8$~nm, $T=300$~K) this yields the thermal oscillation amplitude $\sigma=0.5$~nm. 	
Inserting this into Heisenberg's uncertainty relation, we can derive a minimal slit width that is still compatible with the observation of coherent matter-wave diffraction, for which $\Delta p^\text{grat} > \Delta p^\text{diff}$ holds:
\begin{equation}
s>3.56 \pi\cdot \sigma
\end{equation}
This amounts to $s>5.6$~nm for our nanoscrolls. Such closely spaced nanoscrolls cannot be grown from cutting extended nanoribbons, since these ribbons define their minimal separation. In the future experiment it may become possible to tailor an array of single-wall carbon nanotubes by advanced nanomechanical manipulation techniques. But although it is conceivable to place them as close as $5$~nm, the inter-tube forces will attract the tubes to each other and prevent the formation of a stable grating. It thus seems that for foreseeable future practical nanogratings the momentum recoil will remain within the quantum uncertainty.	

\subsection{Properties of the diffraction gratings}
\label{app:gratings}

\subsubsection{Nanofabrication and characterization of the diffraction gratings}

All diffraction gratings were milled with a focused ion beam (Raith ionline FIB) of $35$~keV Ga$^+$ ions \cite{Lemme2009, Song2011, Wei2010} into the freestanding membranes. Spots with homogeneous membranes were first identified using scanning electron microscopy (SEM, JEOL $6700$, or Raith $150$ ebeam writer, used as SEM). After the writing process all conducting gratings (single and bilayer graphene and the nanoscroll grating) were investigated with transmission electron microscopy (TEM, Delong Instruments LVEM5) or scanning transmission electron microscopy (STEM, Nion UltraSTEM 100). The STEM was operated at $60$~kV with a medium angle annular dark field (MAADF) detector. This inspection is needed to ensure the presence of a clean binary diffraction grating, rather than a patterned heterostructure which may form in single-layer graphene at low ion doses \cite{Kotakoski2015}.
The gratings fabricated from insulating membranes (carbonaceous biphenyl membrane and silicon nitride membrane) were investigated either with environmental scanning electron microscopy (ESEM, FEI Quanta $200$) or electron energy loss spectrometry (EELS, FEI TECNAI F$20$) to determine the parameter of the gratings. All gratings maintained their structure and diffraction properties over time. Virtually identical interferograms can be recorded after storing the gratings for several months in air. 
Some ambiguity in the local thickness of all gratings remains since even though we can confirm the predicted structure and thickness. As for instance shown for single-layer graphene in Fig. \ref{fig:gratings}d, residual atomic and molecular contaminations on the membranes cannot be excluded, as shown in the right lower corner of the same figure. 

In the following the beam parameters for milling of the ultra-thin gratings are given. The dose of the FIB is given in pC/cm instead of pC/cm$^2$. Since we were writing single lines the dose is given along the line direction. Also the step size refers to this direction. The width is defined by the thickness of the beam, which is about $20-30$~nm.

\subsubsection{Carbonaceous biphenyl membrane}

The carbonaceous biphenyl membrane \cite{Angelova2013} is deposited on a copper TEM mesh (Plano $S147-3$) with an additional layer of lacey carbon for stabilization. To mill the grating we used a dose of $12.7$~nC/cm with a dwell time of $0.932$~ms, a current of $2.75$~pA and a step size of $2$~nm. The overall size of the grating in the diffraction experiments is $9.9 \times 5.2$~$\mu$m$^2$. The grating has a periodicity of $107\pm9$~nm, and the grating bars are $54\pm4$~nm wide and 9$77\pm10$~nm long. From the mean width of the slits of $52\pm8$~nm we determine the opening fraction to be $49\pm8$\%. The error bar on the slit width, here and for all other gratings, is the $1\sigma$ standard deviation which was derived from about $100$ individual measurements at different locations of each membrane.
 
\subsubsection{Single layer graphene}
 
The diffraction gratings were written into single-layer graphene (Ted Pella Inc., Redding/CA $21712-5$ PELCO) which was freely suspended over circular holes in a silicon nitride (SiN$_\text{x}$) membrane. Each slit was manufactured by milling two adjacent lines separated by $20$~nm into the membrane. For each line the gallium ion beam was set to a current of $7.1$~pA with a dose of $3$~nC/cm, a step size of $1$~nm and a dwell time of $0.0421$~ms. The grating region has an extension of $7 \times 8$ holes of the support structure, resulting in a size of $30 \times 30$~$\mu$m$^2$. The grating has a periodicity of $101\pm2$~nm. The bars have a mean width of $41\pm6$~nm and are separated by $59\pm6$~nm wide slits, resulting in an opening fraction of $58\pm6$\%. The length of the grating bars is $247\pm8$~nm. As several lines of grating are written on top of each other we have another grating perpendicular to the first one. The periodicity of this second grating is $343\pm10$~nm. Diffraction at this grating parallel to gravity can in principle lead to a stronger mixing of the velocities. However, the periodicity of $343$~nm is too large to result in a visible effect at the detector. 

\subsubsection{A grating of carbon nanoscrolls self-organized from graphene nanoribbons}
\label{app:scrolls}

\begin{figure}[htb]
\begin{center}
\includegraphics[width=8cm]{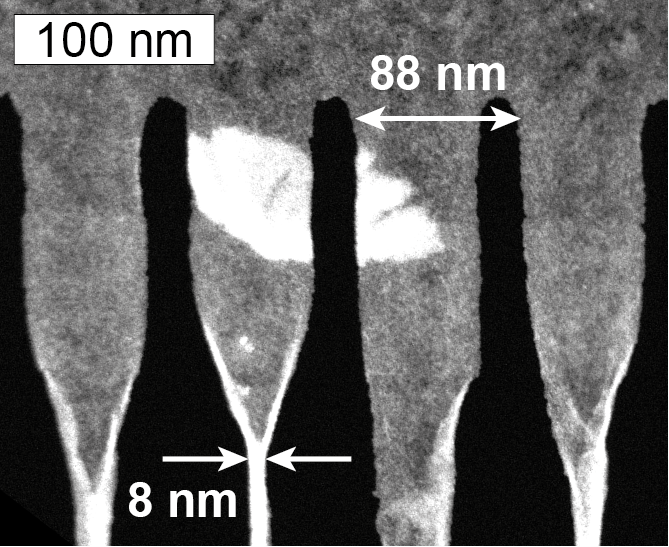}
\end{center}
\caption{Effects determining the opening fraction of the nanoscroll grating. The original width of the grating bars is preserved when foldings in the membrane or contaminations are present. If these factors are not present nanoscrolls are formed with a minimum diameter of down to $8$~nm.}
\label{fig:scrolls}
\end{figure}

For this grating we used the same material as for the single layer graphene grating. The dwell time for FIB milling was $0.232$~ms, the current $6.17$~pA at a step size of $2$~nm and a dose of $7.15$~nC/cm. The grating region used for the diffraction experiments extends over $5.5 \times 18$ holes of the silicon nitride support structure which corresponds to an overall size of $24 \times 73$~$\mu$m$^2$. From parts of the grating which are not rolled up we determined the parameter of the original grating. The bars are $64\pm3$~nm wide, have a length of $1.336\pm3$~$\mu$m and are separated by $23\pm4$~nm wide slits. The periodicity determined is $87\pm5$~nm and the opening fraction is $26\pm6$\%. As expected, the curling does not alter the periodicity in the middle of the grating ($88\pm3$~nm), but increases the opening fraction by almost a factor of $3$ to $74\pm8$\%. The minimum diameter observed in this region is around $8$~nm. Foldings in the membrane as well as contaminations (see Fig. \ref{fig:scrolls}) can disturb this process and lead to a mean width of the grating bars of $23\pm7$~nm. 

\subsubsection{Silicon nitride grating}

The SiN$_\text{x}$ grating has an overall size of $3.3 \times 97$~$\mu$m$^2$. The grating bars are $50\pm2$~nm wide and $956\pm5$~nm long. From the periodicity of $105\pm1$~nm the mean width of the slits results to be $55\pm2$~nm, and the resulting opening fracture is $48\pm3$\%.

\subsubsection{Bilayer graphene grating on lacey carbon}
 
For this sample we used a bilayer graphene sample deposited on a lacey carbon film suspended across a copper TEM mesh (PELCO$^\text{\textregistered}$ $2$-layer Graphene TEM Support Films on Lacey Carbon, $300$ Mesh Copper Grids). The parameters during the milling process were as follows: current $6.03$~pA, step size $2$~nm, dwell time $0.414$~ms and a dose of $12.5$~nC/cm. The resulting grating has a size of $19.5 \times 5.5$~$\mu$m$^2$ and consists of grating bars which are $827\pm3$~nm long and $49\pm8$~nm wide. From the width of the slits ($63\pm7$~nm) and the periodicity of the grating of $112\pm2$~nm, the opening fraction was determined to be $56\pm7$\%.  
For comparison we state the masses of the molecule and a single bar of the diffraction gratings in Table \ref{tab:masses}.  

\begin{table}
\caption{Comparison of the mass of a single grating bar for each grating to the mass of a diffracted molecule (PcH$_2$).}
\begin{tabular}{llcl}
\hline\hline
PcH$_2$	molecule							& m$_\text{PcH$_2$}$ 					& = & $8.5\cdot 10^{-25}$ kg \\ 
Single-layer graphene ribbon	& m$_\text{SLG}$ 							& = & $7.7\cdot 10^{-21}$ kg \\ 
Bilayer graphene ribbon				& m$_\text{BLG}$ 							& = & $6.2\cdot 10^{-20}$ kg \\ 
Carbon nanoscroll							& m$_\text{CNS}$ 							& = & $6.5\cdot 10^{-20}$ kg \\ 
Silicon nitride ribbon				& m$_\text{SiN$_\text{x}$}$ 	& = & $7.5\cdot 10^{-18}$ kg \\ 
Biphenyl membrane ribbon			& m$_\text{BP}$ 							& = & $4.4\cdot 10^{-20}$ kg \\ 
\hline\hline
\end{tabular}
\label{tab:masses}
\end{table}

\subsection{Closing fractures in the ultra-thin membranes}

\begin{figure*}[htb]
\begin{center}
\includegraphics[width=16cm]{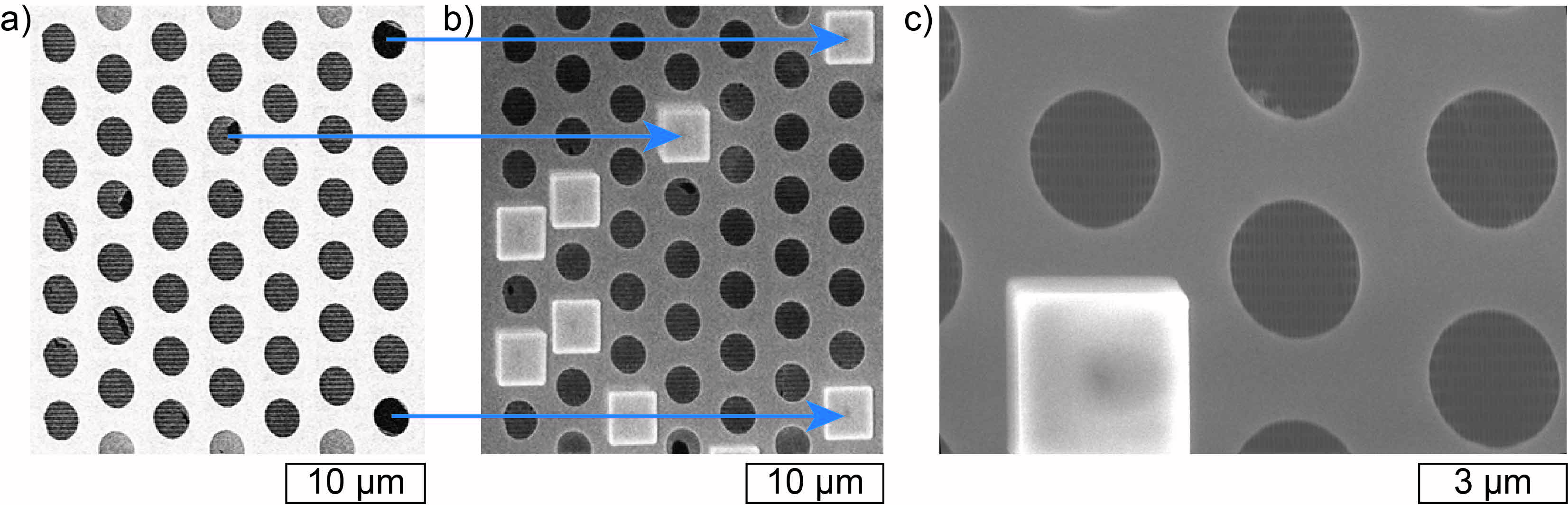}
\end{center}
\caption{Closing grating defects: (a) In the presence of tetraethoxysilane the focused ion beam allows to deposit silicon oxide (SiO$_\text{x}$) and to close open defects. (b) Grating written into a single layer of graphene exhibiting several fractures. After the deposition all larger gaps were closed and only a minimal number of new fractures was introduced. (c) A SEM image at high resolution confirms that the grating is not affected by the deposition process.}
\label{fig:closing}
\end{figure*}

Gratings in ultra-thin membranes may have some defects. These membranes are seldom closed over large areas and may break due to mechanical stress. The resulting fractures in the membrane complicate the analysis of the recorded interferograms as the additional transmission through these holes leads to an artificial enhancement of the zeroth order diffraction peak. We applied FIB also as a method to close these holes even after the grating has been milled, keeping the gratings intact. It allows us to deposit silicon oxide (SiO$_\text{x}$) at the fracture. Figure \ref{fig:closing} shows a single layer graphene grating before (a) and after (b) the treatment. With this method a hole of $3\times3$~$\mu$m$^2$ can be closed within $4$ minutes, which allows manufacturing gratings of higher quality and larger size than before. 

\subsection{Diffraction experiment}

The diffraction experiment is similar to an earlier setup \cite{Juffmann2012}. In short, it 
starts with a $60$~mW continuous laser beam at $421$~nm, which is focused by a $50\times$ 
objective onto the inside of a vacuum window that was coated with phthalocyanine molecules (
PcH$_2$, $m = 514$ atomic mass units, static polarizability \cite{Ramprasad2006} $\alpha = 4
\pi \varepsilon_0 \times (135.7 / 139.9 / 27.5)$~\AA$^3$). The tight laser focus of $1.5$~$\mu
$m defines a transverse coherence angle of $3$~$\mu$rad for molecules traveling at $260$~m/s. They fly about $1.55$~m to the diffraction grating and another $0.6$~m to the detector 
quartz window downstream. The quantum nature of the diffraction is revealed by analyzing 
the position distribution of the arriving particles, which we image using laser-induced 
fluorescence microscopy. 

\subsubsection{Analysis of the diffraction images in Figure \ref{fig:overview}}

Figure \ref{fig:overview} in the main text shows the observed diffraction patterns of PcH$_2$ behind all ultra-thin membranes. We start the analysis by assuming a Maxwell-Boltzmann velocity distribution of the molecular beam. Molecules with different velocities $v$ have different de Broglie wavelengths $\lambda_{dB}=h/mv$, with Planck's constant $h$ and the atomic mass $m$. After two meters of free flight in our vacuum chamber fast molecules land at a higher and slow molecules at a lower screen position. This and the diffraction formula $n\cdot \lambda_{dB}= d\cdot sin⁡(\theta)$ allows calculating the molecular velocity from the position of its interference peaks relative to the zeroth order (see Sec. \ref{sec:velocities}). 

\subsubsection{Quantifying the molecule-surface interaction}
\label{app:Kirchhoff}

\begin{figure*}[htb]
\begin{center}
\includegraphics[width=15cm]{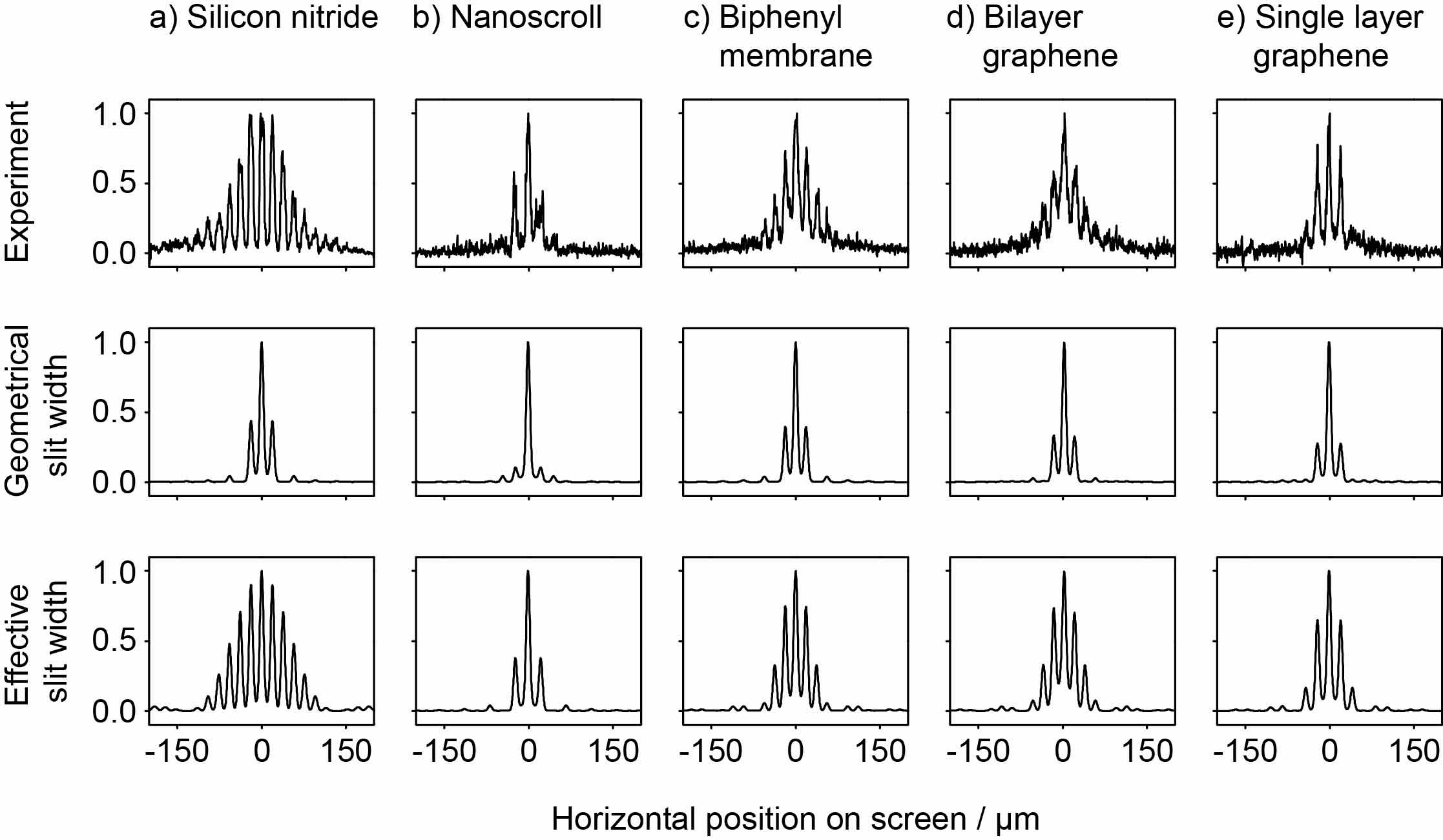}
\end{center}
\caption{Upper row: experimental interference curves, traces as in Figure \ref{fig:overview}. Middle row: Simulation of the interferograms assuming the geometrical slit width. Bottom row: Simulations of the interferograms using an effective slit width that approximates both the amplitude distribution and the number of higher diffraction orders in the experiment. Due to possible scattering, diffusion and vibrations at the detector we convolute the diffraction patterns here with a Gaussian curve with a width of ($\sigma = 3.5$~$\mu$m) \cite{Juffmann2012}. Here we see that the experimental data can be approximated by Kirchhoff-Fresnel diffraction theory if we account for the van der Waals forces \cite{Grisenti1999}:  (a) geometrical slit width $s= 50\pm2$~nm, effective slit width $s_\text{eff} = 15$~nm, (b) $s=65\pm6$~nm, $s_\text{eff} = 49$~nm, (c) $s=54±4$~nm, $s_\text{eff} = 28$~nm, (d) $s=62\pm8$~nm, $s_\text{eff} = 28$~nm, (e) $s=59\pm6$~nm, $s_\text{eff} = 35$~nm.}
\label{fig:fits}
\end{figure*}

Since a full van der Waals description is beyond the scope of this publication, we account for the impact of the molecule-wall interaction by using a reduced effective slit width in the simulation of the interferograms \cite{Bruhl2002, Grisenti1999, Cronin2005, Perreault2005}. Although this simplification cannot give a full description of all observed phenomena, it allows comparing the different gratings. We use a numerical evaluation of the Fresnel- Kirchhoff integral to compute our interference patterns:
\begin{equation}
I_\text{diff} (x^\prime)=I_0 \sum_{n=1}^N{\int_{n∙d-s/2}^{n∙d+s/2}{\exp(ik\sqrt{L^2+(x-x^\prime)^2}) dx}}
\end{equation}
Here $I$ is the observed intensity at a certain position $x^\prime$ at the screen, $N$ is the number of coherently illuminated grating slits, $s$ is the width of each grating slit, $k$ is the wave vector for molecules with the velocity $v$, and $x$ is the coordinate at the grating. The coherence width is determined by the van Cittert-Zernike theorem \cite{Born1993} and two experimentally determined parameters, \emph{i.e.} the source width as well as the distance between the source and the grating.
In Fig. \ref{fig:fits} we compare the results of our experiments (top row, same as Figure \ref{fig:overview} bottom row) with simulations that either take the measured geometrical open slit width (middle row) or an effective slit width $s_\text{eff}$ (bottom row) into account. Reducing the slit width mimics the effect of the van der Waals interactions since both lead to a population of higher diffraction orders. Close to the grating bars, the van der Waals forces can actually become strong enough to deflect the molecules beyond the detector area.   

\subsubsection{The influence of molecular adsorption}

We collect about $20'000$ to $30'000$ molecules per diffraction image. Since the opening fraction of the gratings is about $50$\%, at most the same amount of molecules can stick to the grating. For the single layer graphene sample an open surface area of $49$~$\mu$m$^2$ was illuminated which leads to a maximal deposition density of $6\times 10^{10}$ molecules per cm$^2$. This corresponds to $0.1$\% of the grating surface. Hence, the surface contamination by the investigated molecules can be neglected. In order to minimize the potential partial coverage by molecules of the residual gas, such as nitrogen or water, we kept the gratings in the high vacuum at $p<10^{-7}$~mbar for two weeks before recoding the diffraction images.  

\subsection{Testing de Broglie's relation in the gravitational field}
\label{sec:velocities}

The original pictures recorded with a CCD camera (Andor iXon EMCCD DV8285\_BV) have a size of $1003\times 1004$ pixels. They were background-corrected by recording an image under identical illumination conditions before the deposition of the molecules and by subtracting this image from the interference pattern. As the vertical position of the grating differs slightly between the experiments the absolute position of the interferograms varies also on the detector. Hence, the images in Figure \ref{fig:overview} were shifted and clipped such that they all have the same velocity scale. To determine the velocity, the original images were vertically divided into $20$ stripes. For each of these stripes the velocity was fitted separately. Since the position of the maxima has an uncertainty, so does the experimentally determined velocity at a given height. This is why the data (velocity vs. relative height on the detector) were fitted with the following function to determine the final velocity,
\begin{equation}
v=\sqrt{\frac{g(L\cdot L_1-L^2)/2}{y_2-y_0-(y_1-y_0)\cdot L/L_1}}
\label{eqn:velocity}
\end{equation}
Where $g = 9.81$~m/s$^2$ is the gravitational acceleration, $L=2140$~mm is the distance between the source and the detection window and $L_1=1554$~mm is the distance between the source and the grating. The variable $y$ describes the (unknown but fixed) vertical position of the source ($y_0$) and the grating ($y_1$) as well as the (observed and velocity-dependent) position of the molecules on the detection window ($y_2$). 
\begin{figure}[htb]
\begin{center}
\includegraphics[width=7cm]{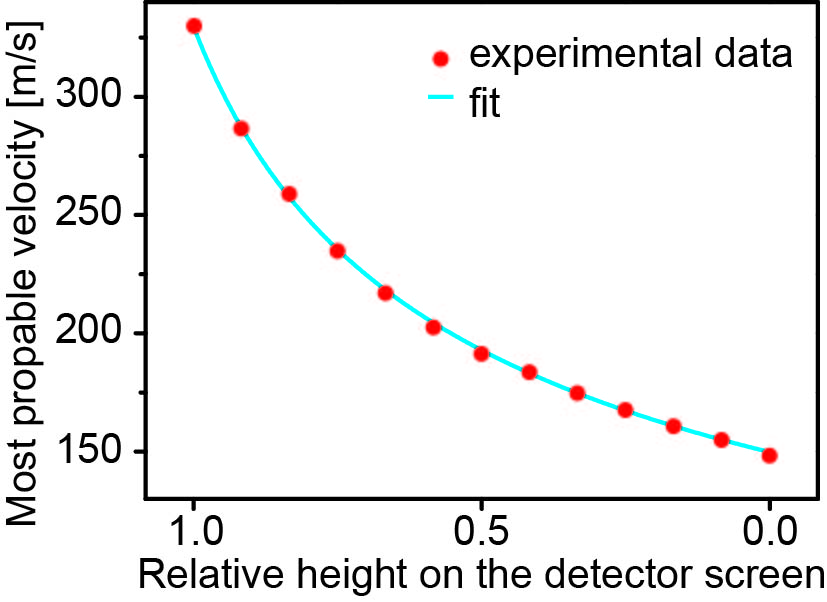}
\end{center}
\caption{Molecular velocity as a function of the vertical detector position. It was fitted with Eq. \ref{eqn:velocity} in discrete positions to verify the agreement between the de Broglie formula $\lambda= h/mv$ and the gravitational free fall curves.}
\label{fig:velocity}
\end{figure}
The result of this fit is shown in Figure \ref{fig:velocity} for the diffraction signal behind the silicon nitride grating. The velocity scale bar given in Figure \ref{fig:overview} is set as a mean for all. Averaging over all five different grating types we can assign the velocities with a standard deviation of $\sigma_v\simeq 12$~m/s for $v= 145$~m/s, and $\sigma_v \simeq 25$~m/s for $v= 263$~m/s

The fit confirms the expected free-fall of all molecules according to Newton's law, or the de Broglie relation for the gravitationally expected velocities. All images of Figure \ref{fig:overview} were vertically stretched by $30$\% for legibility, without any influence on the velocity or color scaling. The color scale bar was also optimized for maximum legibility of the interferograms: black and white correspond to the extremes of $0$ and $100$\% intensity, while the center of the red (yellow) color corresponds to a relative value of $60$\%. The full color map is given in Table \ref{tab:RGB}.
\newpage
\begin{table}[p]
\centering
\caption{Color map "Hot": 24 bit RGB}
\setlength{\tabcolsep}{2mm}
\begin{tabular}{|rrrr|c|rrrr|c|rrrr|}
\hline\hline
Signal	&	R		&		G		&		B	& \qquad\qquad	&	Signal	&	R		&		G		&	B & \qquad\qquad &	Signal	&	R		&		G		&	B \\
\hline													
0.00&		   0  &0			  &0 		&	&0.34& 1 			&0.43457 	&0		& &0.68& 1 			&0.98278 	&0.225\\
0.01&	0.07143 &0			  &0 		&	&0.35& 1 			&0.4563 	&0  	&	&0.69& 1 			&0.98333 	&0.25 			\\
0.02& 0.14286 &0 				&0 		&	&0.36& 1 			&0.47802 	&0  	&	&0.70& 1 			&0.98389 	&0.275	\\
0.03& 0.21429 &0 				&0 		&	&0.37& 1 			&0.49975 	&0  	&	&0.71& 1 			&0.98444 	&0.3  		\\
0.04& 0.28571 &0 				&0 		&	&0.38& 1 			&0.52148 	&0  	&	&0.72& 1 			&0.985 		&0.325		\\
0.05& 0.35714 &0 				&0 		&	&0.39& 1 			&0.54321 	&0  	&	&0.73& 1 			&0.98556 	&0.35 			\\
0.06& 0.42857 &0 				&0 		&	&0.40& 1 			&0.56494 	&0 		&	&0.74& 1 			&0.98611 	&0.375		\\
0.07& 0.5 		&0 				&0 		&	&0.41& 1 			&0.58667 	&0  	&	&0.75& 1 			&0.98667 	&0.4  			\\
0.08& 0.57143 &0 				&0 		&	&0.42& 1 			&0.6084 	&0  	&	&0.76& 1 			&0.98722 	&0.425			\\
0.09& 0.64286 &0 				&0 		&	&0.43& 1 			&0.63012 	&0  	&	&0.77& 1 			&0.98778 	&0.45 	 			\\
0.10& 0.71429 &0 				&0 		&	&0.44& 1 			&0.65185 	&0  	&	&0.78& 1 			&0.98833 	&0.475 				\\
0.11& 0.78571 &0 				&0 		&	&0.45& 1 			&0.67358 	&0  	&	&0.79& 1 			&0.98889 	&0.5   			\\
0.12& 0.85714 &0 				&0 		&	&0.46& 1 			&0.69531 	&0  	&	&0.80& 1 			&0.98944 	&0.525  \\
0.13& 0.92857 &0 				&0  	&	&0.47& 1 			&0.71704 	&0  	&	&0.81& 1 			&0.99 		&0.55 	\\
0.14& 1 			&0 				&0  	&	&0.48& 1 			&0.73877 	&0  	&	&0.82& 1 			&0.99056 	&0.575	\\
0.15& 1 			&0.02173 	&0 		&	&0.49& 1 			&0.76049 	&0 		&	&0.83& 1 			&0.99111 	&0.6  	\\
0.16& 1 			&0.04346 	&0  	&	&0.50& 1 			&0.78222 	&0  	&	&0.84& 1 			&0.99167 	&0.625	\\
0.17& 1 			&0.06519 	&0  	&	&0.51& 1 			&0.80395 	&0  	&	&0.85& 1 			&0.99222 	&0.65  	\\
0.18& 1 			&0.08691 	&0  	&	&0.52& 1 			&0.82568 	&0  	&	&0.86& 1 			&0.99278 	&0.675 	\\
0.19& 1 			&0.10864 	&0  	&	&0.53& 1 			&0.84741 	&0  	&	&0.87& 1 			&0.99333 	&0.7  	\\
0.20& 1 			&0.13037 	&0  	&	&0.54& 1 			&0.86914 	&0  	&	&0.88& 1 			&0.99389 	&0.725  \\
0.21& 1 			&0.1521 	&0  	&	&0.55& 1 			&0.89086 	&0  	&	&0.89& 1 			&0.99444 	&0.75  	\\
0.22& 1 			&0.17383 	&0  	&	&0.56& 1 			&0.91259 	&0  	&	&0.90& 1 			&0.995 		&0.775  \\
0.23& 1 			&0.19556 	&0  	&	&0.57& 1 			&0.93432 	&0  	&	&0.91& 1 			&0.99556 	&0.8  	\\
0.24& 1 			&0.21728 	&0  	&	&0.58& 1 			&0.95605 	&0  	&	&0.92& 1 			&0.99611 	&0.825  \\
0.25& 1 			&0.23901 	&0  	&	&0.59& 1 			&0.97778 	&0  	&	&0.93& 1 			&0.99667 	&0.85  	\\
0.26& 1 			&0.26074 	&0  	&	&0.60& 1 			&0.97833 	&0.025&	&0.94& 1 			&0.99722 	&0.875  \\
0.27& 1 			&0.28247 	&0  	&	&0.61& 1 			&0.97889 	&0.05 &	&0.95& 1 			&0.99778 	&0.9  	\\
0.28& 1 			&0.3042 	&0  	&	&0.62& 1 			&0.97944 	&0.075&	&0.96& 1 			&0.99833 	&0.925  \\
0.29& 1 			&0.32593 	&0  	&	&0.63& 1 			&0.98 		&0.1  &	&0.97& 1 			&0.99889 	&0.95  	 \\
0.30& 1 			&0.34765 	&0  	&	&0.64& 1 			&0.98056 	&0.125&	&0.98& 1 			&0.99944 	&0.975  \\
0.31& 1 			&0.36938 	&0  	&	&0.65& 1 			&0.98111 	&0.15 &	&0.99& 1 			&1 				&1 			\\
0.32& 1 			&0.39111 	&0  	&	&0.66& 1 			&0.98167 	&0.175&	&		 &				&					&				\\
0.33& 1 			&0.41284 	&0  	&	&0.67& 1 			&0.98222 	&0.2  &	&		 &			  &					&				\\	
\hline\hline
\end{tabular}
\label{tab:RGB}
\end{table}

\end{document}